\documentclass{lmcs}

\newcommand{\bull}{\rule{.85ex}{1ex} \par \bigskip}

\newtheorem{theorem}{Theorem}[section]

\newtheorem{proposition}[theorem]{Proposition}
\newtheorem{lemma}[theorem]{Lemma}
\newtheorem{corollary}[theorem]{Corollary}

\newcommand{\cal}[1]{\ensuremath{\mathcal{#1}}}

\usepackage{cite}
\usepackage[latin1]{inputenc}
\usepackage{amsmath, amsxtra, amsfonts, amssymb}
\usepackage{graphicx}
\usepackage{cite}
\usepackage{picin/picins}
\usepackage{xspace, varioref}
\usepackage[matrix, curve, arrow, tips, frame]{xy}

\begin{document}

\title[Horn versus full first-order]{Horn versus full first-order: complexity dichotomies in algebraic  
constraint satisfaction}

\author{Manuel Bodirsky}
\address{CNRS/LIX, \'Ecole Polytechnique, France}
\email{bodirsky@lix.polytechnique.fr}
\author{Peter Jonsson}
\address{Department of Computer and System Science, Link\"opings Universitet, Link\"oping, Sweden}
 \email{petej@ida.liu.se}
\author{Timo von Oertzen}
\address{Max-Planck-Institute for Human Development, Berlin, Germany}
\email{vonoertzen@mpib-berlin.mpg.de}

\maketitle
\thispagestyle{empty}

\begin{abstract}
We study techniques for deciding the computational complexity
of infinite-domain constraint satisfaction problems. 
For certain fundamental algebraic structures $\Delta$, we prove
definability dichotomy theorems of the following form:
for every first-order expansion $\Gamma$ of $\Delta$,
either $\Gamma$ has a quantifier-free Horn definition in $\Delta$, or
there is an element $d$ of $\Gamma$ such that 
all non-empty relations in $\Gamma$ contain a tuple of the form $(d,\dots,d)$,  or all relations with a first-order definition in $\Delta$
have a primitive positive definition in $\Gamma$. 

The results imply that several families of constraint satisfaction
problems exhibit a complexity dichotomy: the problems are in P
or NP-hard, depending on the choice of the allowed relations. 
As concrete examples, we investigate fundamental algebraic constraint satisfaction problems. The first class consists of 
all first-order expansions of $(\mathbb Q;+)$.
The second
class is the affine variant of the first class. In both cases, we obtain
full dichotomies by utilising our general methods.
\end{abstract}

\section{Introduction}
Constraint satisfaction problems (CSPs) are computational problems
that appear in almost every area of computer science such as 
artificial intelligence, graph algorithms, scheduling, combinatorics, and computer algebra. 
Depending on the type of constraints that are allowed in the input instances of a CSP, the
computational complexity of a CSP is usually polynomial (we will call
these CSPs tractable), or NP-hard. In the last decade, a lot of progress
was made to find general criteria that imply that a CSP is tractable,
or that it is NP-hard. 
Such results have been obtained for constraint languages over finite domains~\cite{JBK,Maltsev,IMMVW,BulatovLICS}, but also 
for constraint languages over infinite domains that are $\omega$-categorical (for formal definition of these concepts see Section~\ref{sect:csp}). For example, it has been shown that for every structure $\Gamma$ with a first-order
definition in $(\mathbb Q;<)$ the problem CSP$(\Gamma)$
is in P if it falls into one out of nine classes, and is NP-hard otherwise~\cite{tcsps}.

Lately, many researchers have been fascinated by a 
conjecture due to Feder
and Vardi~\cite{FederVardi} which is known as the \emph{dichotomy conjecture}.
This conjecture says that every CSP with a finite
domain constraint language is either tractable (i.e., in P) or NP-complete. According to
a well-known result by Ladner~\cite{Ladner}, there are
NP-\emph{intermediate} computational problems, i.e., problems in NP that are
neither tractable nor NP-complete (unless P=NP). But the problems that are given in Ladner's construction are extremely artificial.  The question why
there are so few candidates for natural NP-intermediate problems is one of the mysteries in complexity theory.  

Any outcome of the dichotomy conjecture is
probably surprising: a negative answer would finally provide
relatively natural NP-intermediate problems, which would be of 
interest in complexity theory. A positive answer probably comes
with a criterion which describes the NP-hard CSPs (and it would
probably even provide algorithms for the tractable CSPs). But then we
would have a rich catalogue of computational problems
where the computational complexity is known. Such a catalogue would be a valuable tool for deciding the complexity of computational problems
in the mentioned application areas: since CSPs are abundant, one might
derive algorithmic results by reducing the problem of interest to a
known tractable CSP, and one might derive hardness results by reducing a known NP-hard CSP to the problem of interest.


In this article, we study two natural classes of infinite domain constraint languages, and show that the corresponding CSPs do exhibit a complexity dichotomy. To the best of our knowledge,
this is the first systematic complexity result for classes of structures that are not $\omega$-categorical.
The first class consists of all first-order expansions of 
$(\mathbb Q; \{(x,y,z) \; | \; x+y=z\})$ (i.e., we add relations to $(\mathbb Q; \{(x,y,z) \; | \; x+y=z\})$ that are first-order definable in $(\mathbb Q; \{(x,y,z) \; | \; x+y=z\})$.
The second class is an affine version of the first class,
and consists of all first-order expansions of 
$(\mathbb Q; \{(a,b,c,d) \; | \; a-b+c = d\})$.
That the structures $(\mathbb Q; \{(x,y,z) \; | \; x+y=z\})$ and $(\mathbb Q; \{(a,b,c,d) \; | \; a-b+c = d\})$ are not $\omega$-categorical follows
immediately from the theorem by Engeler, Ryll-Nardzewski, and 
Svenonius (cf. Theorem 6.3.1 in \cite{Hodges}). 
It is even the case that the corresponding CSPs cannot be
formulated by any $\omega$-categorical template; the basic proof idea
is presented in \cite[Proposition 1]{BodirskySurvey}; also see~\cite{BodHilsMartin}.

Our results follow from theorems about primitive positive
definability: we show that 
for every relation $R$ with a first-order definition in $({\mathbb Q};+)$,
either $R$ has a quantifier-free Horn definition in $({\mathbb Q};+)$, or 
$R$ contains the tuple $(0,\dots,0)$, or
all  relations with a first-order definition in $({\mathbb Q};+)$
have a primitive positive definition in $({\mathbb Q};+,R)$.
The analogous result also holds for the affine case.
The techniques that we use to prove these two definability theorems 
are more general than the two 
classification results, and they are very different in nature.
One technique applies for structures `that have little
structure'; to be precise, for \emph{all} structures $\Gamma$ 
where $=$ and $\neq$ are the only primitive positive definable non-trivial binary relations (Section~\ref{sect:affine}). In particular,
they apply to structures with a 2-transitive automorphism group.
The other technique applies for structures `with a lot of structure';
informally, it applies whenever we can find a primitive positive definition for the line between two
points in ${\mathbb Q}^k$ (Section~\ref{sect:q}).

The rest of this paper is organised as follows: in Section~\ref{sect:csp}, we provide
some background material on constraint satisfaction and logic.
A tractability result for templates that have a quantifier-free Horn definition in
$(\mathbb Q;+)$ is presented in Section~\ref{sec:tract}.
The classification result for $(\mathbb Q;+)$ can be found in
Section~\ref{sect:q} while the results for the affine case are collected in Section~\ref{sect:affine}. 
Finally, a number of open questions and directions for
future work can be found in Section~\ref{sect:conc}.

\section{Preliminaries}
\label{sect:csp}
Let $\Gamma=(D; R_1,\dots,R_n)$ be a relational structure\footnote{Our terminology is standard; all notions that are not introduced 
in the article can be found in standard text books, e.g., in~\cite{Hodges}.} 
with domain $D$ (which will usually be infinite) and finitely many relations $R_1,\ldots,R_n$.
The \emph{constraint satisfaction problem for $\Gamma$} 
(short, CSP$(\Gamma)$)
is the computational problem to decide whether a given 
primitive positive sentence $\Phi$ involving relation
symbols for the relations in $\Gamma$ is true in $\Gamma$.
A first-order formula is called \emph{primitive positive}
if it is of the form 
\begin{align*}
\exists x_1,\dots,x_n. \psi_1 \wedge \ldots \wedge \psi_m
\end{align*}
where $\psi_i$ are atomic formulas, i.e., formulas of the form
$x=y$ or $R(x_{i_1},\dots,x_{i_k})$ with $R$ the relation
symbol for a $k$-ary relation from $\Gamma$. We call such
a formula a {\em pp-formula}.
The conjuncts in a pp-formula $\Phi$ are also called the
\emph{constraints} of $\Phi$.
We also refer to $\Gamma$ as
a \emph{constraint language} (it is also often called the \emph{template}) of CSP$(\Gamma)$.

We say that a first-formula $\phi$ \emph{defines}
a relation $R$ in $\Gamma$ when
$\phi(a_1,\dots,a_k)$ holds in $\Gamma$ iff $(a_1,\dots,a_k) \in R$.
If $\phi$ is primitive positive, we call $R$ \emph{primitive positive definable (pp-definable)} over $\Gamma$.
The following simple but important result explains the importance
of primitive positive definability for constraint satisfaction problems.

\begin{lemma} \label{pp-def}
Let $\Gamma$ be a relational structure and $\Gamma'$ be an
expansion of this structure by a pp-definable relation $R$ over $\Gamma$. Then CSP$(\Gamma)$ is polynomial-time
equivalent to CSP$(\Gamma')$.
\end{lemma} 

Lemma~\ref{pp-def} will be used extensively in the sequel and we
will not make explicit references to it.
Another important class of formulas are 
\emph{Horn} formulas; a first-order formula in conjunctive normal form is Horn if and only if
each clause contains at most one positive literal.
A relation $R$ is called quantifier-free Horn definable over $\Gamma$
if there exists a quantifier-free Horn formula that defines $R$ in $\Gamma$.
Note that Lemma~\ref{pp-def} does not
hold if we replace `pp-definable' with `Horn definable'.

By choosing an appropriate structure $\Gamma$
many computational problems that have been studied in the literature can be formulated as CSP$(\Gamma)$
(see e.g.~\cite{JBK,CSPSurvey,BodirskySurvey}).
It turns out very often that the structure $\Gamma$ can be chosen
to be $\omega$-categorical. A structure is called \emph{$\omega$-categorical} if the set of all first-order sentences that is true in the structure has only one countable model, up to isomorphism.
A famous example of an $\omega$-categorical structure is $(\mathbb Q; <)$. The condition of $\omega$-categoricity  is interesting for constraint satisfaction, because
the so-called \emph{universal-algebraic approach}, which 
is currently intensively studied for finite constraint languages, applies---at least in principle---also for $\omega$-categorical structures (see e.g.~\cite{tcsps} for an application of the universal-algebraic approach
to CSPs for constraint languages over infinite domains).
In this article, we demonstrate that systematic complexity
classification can be performed for constraint languages
over infinite domains even if the constraint languages are not $\omega$-categorical.

\paragraph{\bf Example.}
Let $\Gamma$ denote the structure 
\[(\mathbb Q; \{(x,y,u,v) \; | \; (x=2y \; \vee \; y = u+v ) \wedge x \neq u \})\] 
It can be shown that CSP$(\Gamma)$ \emph{cannot} be
formulated with an $\omega$-categorical template (for a very similar proof, see~\cite{BodirskySurvey}; a necessary and sufficient condition about which CSPs can be formulated with $\omega$-categorical templates can be found in~\cite{BodHilsMartin}). One can show that the relations
$\{(x,y) \; | \; x \neq y\}$ and
$\{(x,y,z) \; | \; x = y+z\}$ have pp-definitions in $\Gamma$. It is now straightforward 
to determine the computational complexity of CSP$(\Gamma)$
by combining Lemma~\ref{pp-def} and our
classification result (Corollary~\ref{cor:q}).

We will sometimes consider the automorphism group Aut$(\Gamma)$ of a template $\Gamma$ over a domain $D$, i.e., the group
formed by the set of all automorphisms\footnote{Isomorphisms between $\Gamma$ and $\Gamma$.} of $\Gamma$ with respect to functional composition.
An \emph{orbit} of $Aut(\Gamma)$ on $D^2$ 
is a set of the form 
$\{(\alpha(a),\alpha(b)) \; | \; \alpha \in Aut(\Gamma)\}$, 
for some $a,b \in D$. 
We note that pairs from the same orbit satisfy the same first-order formulas.

Let $D$ be an arbitrary infinite set and arbitrarily choose an
element $d \in D$.
The complexity of CSP$(\Gamma)$ where $\Gamma$ has
a first-order definition in $(D;=)$ (so-called {\em equality languages}) has been classified in~\cite{ecsps}. We note that if $R$ is first-order definable in $(D;=)$ and $(d,\ldots,d) \in R$, then
$(d',\ldots,d') \in R$ for every $d' \in D$. Thus, the exact choice of $d$
is irrelevant when stating the following theorem.

\begin{theorem}[of~\cite{ecsps}]\label{thm:ecsp}
Let $\Gamma$ be a template with a first-order definition in $(D;=)$.
Then, all relations in $\Gamma$ have a quantifier-free Horn definition in $(D;=)$, or all non-empty relations in $\Gamma$ contain the tuple $(d,\dots,d)$, or else every
first-order definable relation in $(D;=)$ has a pp-definition in $\Gamma$. 
In the last case, CSP$(\Gamma)$ is NP-complete.
\end{theorem}

Instead of using Theorem~\ref{thm:ecsp} in its full generality, it will
be sufficient to use a simple corollary.
For any set $D$, the relation $S_D$ denotes the relation
$$\{(x,y,z) \in D^3 \; | \; y \neq z \wedge (x=y \vee x=z)\} \; .$$

\begin{corollary} \label{s-prop}
Let $D$ be an infinite set.
Every first-order definable relation in $(D; =)$ has a pp-definition
in $(D; S_D)$. 
\end{corollary}
\begin{proof}
The relation $S_D$ has a first-order definition in $(D;=)$ and 
does not contain the tuple $(d,d,d)$.
It is easy to verify that
$S_D$ has no quantifier-free Horn definition in $(D;=)$
so every
first-order definable relation in $(D;=)$ has a pp-definition in 
$(D; S_D)$
by Theorem~\ref{thm:ecsp}.
\end{proof}

\section{Tractability} \label{sec:tract}
For all relational structures $\Gamma$ with  a quantifier-free Horn definition in $(\mathbb Q;+)$, the problem CSP$(\Gamma)$ can
be solved in polynomial time. This follows from a more general
algorithmic result in~\cite{JonssonBaeckstroem}. However, the algorithm presented 
there solves a linear number of linear programs, and
thus the best known algorithms have a rather high worst-case 
running time. We present a more efficient algorithm for the
special case that is relevant in our paper.
We denote by $O^{\sim}(f(N))$ the class of all functions of asymptotic growth
at most $f(N)$ up to poly-logarithmic factors.

\begin{proposition}\label{prop:tract}
Let $\Gamma$ be a relational structure whose
relations have a quantifier-free Horn definition in $(\mathbb Q;+)$.
Then there is an algorithm that solves CSP$(\Gamma)$ in time $O^{\sim}(N^4)$ where $N$ is the size of the input.
\end{proposition}

The algorithm we present in the proof of Proposition~\ref{prop:tract}  is a combination of general techniques in constraint
satisfaction~\cite{Disj,Maximal} 
and a polynomial implementation of Gaussian elimination algorithm on rational data. Since the input of CSP$(\Gamma)$ 
consists of a primitive positive sentence whose atomic formulas
are of the form $R(x_1,\dots,x_k)$ where $R$ is quantifier-free Horn definable over $(\mathbb Q;+)$, we can as well assume that the input
to our problem consists of a set of Horn clauses over $(\mathbb Q;+)$.

We have to make some remarks about the worst-case running time of the Gaussian elimination algorithm. It is well-known that the Gaussian elimination requires $O(n^2m)$ many arithmetic operations on rational numbers, where $m$ is the number of equations and $n$ is the number of variables.
In our algorithm, we have to solve a linear
number of linear equation systems $S_1,\dots,S_m$; however, system $S_{i+1}$ is obtained from system $S_{i}$ by adding a single linear equation. Since the Gaussian algorithm
can be presented in such a way that it computes a system in triangular form, adding successively equation by equation, the overall costs for solving
$S_1,\dots,S_m$ equals the cost to solve $S_m$ with Gaussian elimination.

Also recall that the size of the numbers involved when performing the Gaussian elimination algorithm might grow exponentially when implemented without care.
However, when we use the Euclidean algorithm to shorten the coefficients during
the elimination process, the Gaussian elimination algorithm can be shown to be polynomial~\cite{Edmonds}.
We are only interested
in deciding solvability of linear equation systems, and not constructing solutions, and so we even have \emph{linear} bounds (in the input size) on the representation size of all numbers involved in deciding solvability for linear equation systems over the rational numbers with Gaussian elimination (see~\cite{Schrijver}, proof of Theorem 3.3).
Finally we remark that the most costly arithmetic operation that has to be
performed on rational numbers during the elimination process is multiplication,
and multiplication can be performed in time $O(s \log s \log \log s)$, where $s$ denotes the representation size of the two rational numbers (in bits).
Hence, the overal running time for solving $S_1,\dots,S_m$ with the discussed implementation of the Gaussian elimination algorithm is in $O^{\sim}(N^4)$.

We will show that our algorithm for CSP$(\Gamma)$ can be implemented such that it has the same overall asymptotic worst-case complexity.


\begin{figure}[h]
\begin{center}
\small
\fbox{
\begin{tabular}{l}
Solve($\Phi$) \\
{\rm // Input: An instance $\Phi$ of CSP$(\Gamma)$} \\
{\rm // where all relations in $\Gamma$ have a quantifier-free Horn definition in $(\mathbb Q;+)$} \\
{\rm // Output: \emph{yes} if $\Phi$ is true in $\Gamma$, \emph{false} otherwise} \\
Let $\cal C$ be the set of all Horn-clauses from each constraint in $\Phi$ \\
Let $\cal U$ be the subset of $\cal C$ that only contains clauses with a single positive literal. \\
Do \\
\hspace{.5cm}    For all negative literals $\neg \phi$ in clauses from $\cal C$ \\
\hspace{1cm}    If $\cal U$ implies $\phi$ delete the negative literal $\neg \phi$ from all clauses in $\cal C$. \\
\hspace{.5cm}    If $\cal C$ contains an empty clause then return \emph{unsatisfiable}. \\
\hspace{.5cm}    If $\cal C$ contains a clause with a single positive literal $\psi$, add $\{\psi\}$ to $\cal U$. \\
Loop until no literal has been deleted \\
Return \emph{satisfiable}.
\end{tabular}}
\end{center}
\caption{An algorithm for the constraint satisfaction problem where
all constraint relations have a quantifier-free Horn definition in $(\mathbb Q;+)$.}
\label{fig:alg}
\end{figure}

\begin{proof}[Proof of Proposition~\ref{prop:tract}]
We first discuss the correctness of the algorithm shown in 
Figure~\ref{fig:alg}, and then explain how to implement the algorithm such that it achieves the
desired running time.

When $\cal U$ logically implies $\phi$
then the negative literal $\neg \phi$ is never satisfied and can be deleted from all clauses without affecting the set of solutions.
Since this is the only way how literals can be deleted from clauses, it is clear that if one clause becomes empty the instance
is unsatisfiable. 

If the algorithm terminates with \emph{yes}, then no negation of a disequality
is implied by $\cal U$. If $r$ is the rank of the linear equation system defined by $\cal U$, 
we can use the Gaussian elimination algorithm as described above to eliminate from all literals in the remaining clauses $r$ of the variables. 
Let $S$ be the maximal sum of the absolute values of all coefficients
in one of the remaining inequalities plus one. Then setting the $i$-th
variable to $S^i$ satisfies all clauses. 

To see this, take any disequality, and assume that $i$ is the highest variable
index in this disequality. Order the disequality in such a way that the variable
with highest index is on one side and all other on the other side of the
$\neq$ sign. The absolute value on the side with the $i$-th variable is at
least $S^i$. The absolute value on the other side is less than $S^i - S$, since all
variables have absolute value less than $S^{i-1}$ and the sum of all
coefficients is less than $S-1$ in absolute value. Hence, both sides of the
disequality have different absolute value, and the disequality is satisfied.
Since all remaining clauses have at least one disequality, all
constraints are satisfied.

We finally explain how to implement the algorithm such that it
runs in time $O^{\sim}(N^4)$. To decide whether $\cal U$ implies an
equality $\phi$, we compute in each interation of the main loop
the triangular normal form for the linear equation system
determined by $\cal U$ as described before the statement of the Proposition. The overall costs to do this are in $O^{\sim}(N^4)$.
Moreover, for each negative literal we maintain an equation
where we eliminate as many variables as possible using the computed
triangular normal form. If one of these equations becomes 
trivial (i.e. is the form $a=a$) we conclude that the equation is implied
by $\cal U$. The overall costs for doing this is also bounded by $O^{\sim}(N^4)$ by a very similar argument as given before the statement of the proposition. With appropriate straightforward data structures, the total costs for removing negated literals $\neg \phi$ 
from all clauses when $\phi$ is implied by $\cal U$ is linearly bounded in the input size since each literal can be removed at most once. 
\end{proof}

\section{The Rational Numbers with Addition} \label{sect:q}
In this section we present the complexity classification for 
first-order expansions of 
$(\mathbb Q; \{(x,y,z) \; | \; x+y=z \})$.
We begin in Section~\ref{defneq} with 
a result about the pp-definability of the disequality relation $\neq$
in first-order expansions of $(\mathbb Q; \{(x,y,z) \; | \; x+y=z \})$.
When the relation $\neq$ is pp-definable, we show
that also the relation $S_{\mathbb Q}$ (defined in Section~\ref{sect:csp} as the relation $\{(x,y,z) \in \mathbb Q^3 \; | \;  x \neq z \wedge (x=y \vee y=z) \}$) is pp-definable
whenever the constraint language contains a relation $R$ that is
first-order, but not quantifier-free Horn definable in $({\mathbb Q};+)$;
this is shown in Section~\ref{sect:defsq}.
Finally, Section~\ref{sect:cc1} completes the classification for
first-order expansions of $(\mathbb Q; \{(x,y,z) \; | \; x+y=z \})$.

\subsection{Definability of Disequality}\label{defneq}

\begin{lemma}\label{lem:trans}
For any structure $\Gamma$ with a first-order definition in $(\mathbb Q; +)$, the first-order definable relations in $\Gamma$ are a subset of $\{{\mathbb Q}, {\mathbb Q} \setminus \{0\}, \{0\},\emptyset\}$. 
\end{lemma}
\begin{proof}
Let $R$ be a unary relation with a first-order definition 
in $(\mathbb Q; +)$. The statement is clear if $R$ does not contain any element distinct from $0$, so let $a$ be from ${\mathbb Q} \setminus \{0\}$. We have to show that $R={\mathbb Q}$ or $R=\mathbb Q \setminus\{0\}$. Observe that for any $c \in \mathbb Q$, $c \neq 0$, the mapping $x \mapsto cx$ is an automorphism of $\Gamma$. Hence, 
for any $b \neq 0$
there is an automorphism of $(\mathbb Q; +)$ that maps $a$ to $b$. Since
automorphisms preserve first-order formulas, so $b \in R$ and the claim
follows.
\end{proof}

Note that $x=0$ is equivalent to $x+x=x$ and hence the relation $\{0\}$ is pp-definable over $(\mathbb Q; +)$; thus we can use $0$ freely as
a constant symbol in pp-definitions over $\Gamma$.
\begin{proposition}\label{prop:neq}
Let $\Gamma$ be a first-order expansion 
of $({\mathbb Q}; +)$
containing a non-empty relation $R$ such that $R(x,\dots,x)$ is false for any $x$. Then $\neq$ is pp-definable in $\Gamma$.
\end{proposition}
\begin{proof}
Observe that if the set ${\mathbb Q} \setminus \{0\}$ has a pp-definition $\phi(u)$ in $\Gamma$, then the pp-formula
$$ \exists u,y'. \; \phi(u) \wedge y+y'=0 \wedge x+y'=u $$
defines $x \neq y$ over $\Gamma$.

Let $S$ be a non-empty pp-definable relation in $\Gamma$ of minimal
arity such that $S(x,\dots,x)$ defines the empty set.
Let $k$ be the arity of $S$. 
First, assume that $S(x_1,x_2,\dots,x_k) \wedge x_1=x_2$
is satisfiable. Then the $(k-1)$-ary relation $S'(x_2,\dots,x_k)$ defined
by $S(x_2,x_2,\dots,x_k)$ is non-empty, and $S'(x,\dots,x)$ defines
the empty set; this is in contradiction to the choice of $S$.

Assume next that $S(x_1,\dots,x_k) \wedge x_1=x_2$ is unsatisfiable.
Define the unary relation $T(x)$ by
$$\exists x_3,\dots,x_k. \, S(x,0,x_{3},\dots,x_k)$$
and the unary relation $U(y)$ by 
$$\exists x_1,x_3,\dots,x_k. \, S(x_1,y,x_3,\dots,x_k)\; . $$ 
By Lemma~\ref{lem:trans}, both $T$ and $U$ are from $\{{\mathbb Q}, {\mathbb Q} \setminus \{0\}, \{0\},\emptyset\}$. 
The relation $T$ cannot be equal to $\{0\}$ or to $\mathbb Q$ 
since this contradicts the assumption that $S(x_1,x_2,\dots,x_k) \wedge x_1=x_2$ is unsatisfiable. 
If $T$ is equal to ${\mathbb Q} \setminus \{0\}$, then by the initial observation $\neq$ is pp-definable in $\Gamma$ and we are done.
We conclude that $T = \emptyset$ and hence $0 \notin U$.
Since $U$ is non-empty, it must be the case that $U = \mathbb Q \setminus \{0\}$, and again by the initial observation $\neq$ is pp-definable in $\Gamma$.
\end{proof}

\subsection{Definability of $S_{\mathbb Q}$} \label{sect:defsq}
The rational numbers with addition (and also the real numbers with addition) admit \emph{quantifier elimination}, i.e., every relation with a first-order
definition in $({\mathbb Q};+)$ also has a quantifier-free 
definition over $({\mathbb Q};+)$. This follows from the more general
fact that the first-order theory of torsion-free divisible abelian groups 
admits quantifier elimination (see e.g.~Theorem~3.1.9 in~\cite{Marker}).

The first lemma allows us to freely use certain expressions in pp-definitions over 
$(\mathbb Q; +)$. 

\begin{lemma} \label{lem:pp}
The relation $\{(x_1,\ldots,x_l) \; | \; r_1x_1 + \ldots + r_l x_l = 0\}$ is
pp-definable in $(\mathbb Q; +)$ for arbitrary $r_1,\ldots,r_l \in {\mathbb Q}$.
\end{lemma}

\begin{proof}
First observe that we can assume that $r_1,\dots,r_l$ are integers,
because we can multiply the equation $r_1 x_1 + \dots + r_l x_l = 0$
by the least common multiple of the denominators of $r_1, \dots, r_l$
and obtain an equivalent equation.
The proof is by induction on $l$. We first consider that case that
$l=1$. If $r_1=0$, there is nothing to show.
Otherwise, the formula $r_1 x_1 = 0$ is equivalent to
$x_1 + x_1 = x_1$. Hence, we can in particular use expressions of the
form $x=0$ and $x+y=0$ in pp-definitions over $(\mathbb Q;+)$  with variables $x,y$.
If $l=2$, and $r_1=0$ or $r_2=0$, then we can argue as in the case $l=1$. If $r_1$ and $r_2$ are both positive or both negative, then
$r_1 x_1 + r_2 x_2 = 0$ is equivalent to
\begin{align*}
\exists u_1,\dots, u_{r_1}, v_1,\dots,v_{r_2}. & u_1=x_1 \wedge v_1=x_2 \wedge u_{r_1}+v_{r_2}=0 \; \wedge \\
& \bigwedge_{i=1}^{r_1-1}  x_1 + u_i =  
u_{i+1} \; \wedge \;  \bigwedge_{i=1}^{r_2-1}  x_2 + v_i =  
v_{i+1}  \; .
\end{align*}
If $r_1$ and $r_2$ have different signs, we replace the conjunct
$u_{r_1}+v_{r_2}=0$ in the formula above by $u_{r_1}=v_{r_2}$.

Now suppose that $l>2$. By the inductive assumption,
there is a pp-definition $\phi_1$ for
$r_1x_1 + r_2 x_2 + u = 0$ and a pp-definition $\phi_2$ for $r_3x_3+\dots r_l x_l + v= 0$.
Then $\exists u,v. \phi_1 \wedge \phi_2 \wedge u+v=0$
is a pp-definition for $r_1 x_1 + \dots + r_l x_l = 0$.
 \end{proof}

In the following, $R$ denotes a relation with a quantifier-free first-order definition $\phi$
in $({\mathbb Q};+)$. 
A quantifier-free first-order formula $\phi$ in  
conjunctive normal form is called \emph{reduced} if every formula 
obtained from $\phi$ by removing 
a literal is not equivalent to $\phi$ (this concept was introduced in~\cite{BodChenPinsker}). Clearly, 
such a reduced definition of $R$ always exists, 
because we can find one by successively removing literals from $\phi$.
Note that if $l$ is a literal from $\phi$, 
then $\neg l$ can be written as a pp-formula
over a structure that contains $\neq$ and $+$.

\begin{lemma} \label{lem:hard}
If $R$ is first-order, but not quantifier-free Horn definable in $({\mathbb Q};+)$,
then $S_{\mathbb Q}$
has a pp-definition in $({\mathbb Q}; R,+,\neq)$.
\end{lemma}
\begin{proof}
Let $T(x,y) \subsetneq {\mathbb Q}^2$ be the 
binary relation defined by $x \neq 0 \wedge (y=0 \vee x=y)$. 
We first prove that $T$
has a pp-definition in $({\mathbb Q}; R,+,\neq)$.
Let $\phi$ be a 
reduced first-order definition of $R$, and
let $C$ be a clause of $\phi$ with two positive literals $l_1$ and $l_2$.
Because $\phi$ is reduced, there are 
$p,q \in R$ such that $p$ satisfies $l_1$ and does not satisfy
all other literals in $C$,
and $q$ satisfies $l_2$ but does not satisfy all other literals in $C$.

We claim that the following pp-formula is
logically equivalent to $x \neq 0 \wedge (y=0 \vee x=y)$.
\begin{align*}
\exists z_1,\dots,z_k. \quad  
& x \neq 0 \quad \wedge \quad \bigwedge_{i=1}^k z_i = p_i x + (q_i-p_i) y \quad
\wedge \quad \\ & 
\bigwedge_{l \in C \setminus \{l_1,l_2\}} \neg l  \quad \wedge \quad  R(z_1,\dots,z_k) 
\end{align*}
Let $x \neq 0$ be arbitrary.
Suppose that $y=0$. Then the assignment 
$z_1 = p_1 x, \dots, z_k = p_k x$ obviously satisfies the first line in the pp-formula. Recall that $p \in R$ and 
$p$ does not satisfy all literals in $C$ except for $l_1$.
The function $f(a) =  x \cdot a$ is in Aut$({\mathbb Q};+)$ whenever $x \neq 0$.
Consequently, $f \in \text{Aut}(\mathbb Q; R)$, too, and the second line in the formula is satisfied as well.
Now suppose that $x=y$. Then the assignment $z_1 = q_1 x, \dots, z_k = q_k x$
obviously satisfies the first line in the pp-formula.
By construction, $q \in R$ and $q$ does not satisfy all literals in $C$ except for $l_1$. Again we conclude that the second line in the formula
is also satisfied.

For the opposite direction, suppose that $x,y \in \mathbb Q$ 
satisfy the pp-formula. Because of the first line
of the formula, $x \neq 0$.
Let $z_1,\dots,z_k$ be the
$k$ elements whose existence is asserted in the first line of the formula.
Note that the equations of the first line imply that $(z_1,\dots,z_k)$ lies on the line $L \subset \mathbb Q^k$ defined by $p x$ and $q x$. 
Because the formula contains the conjunct $R(z_1,\dots,z_k)$, the
clause $C$ in $\phi$ is satisfied by $z_1,\dots,z_k$. Since
$z_1,\dots,z_k$ also satisfies 
the conjunction of all negated literals in $C$ except for
the positive literals $l_1$ and $l_2$, at least one of these two literals 
$l_1$ and $l_2$ must be satisfied by $z_1,\dots,z_k$. 

Suppose first that $l_1$ is satisfied.
The line $L$ does not lie completely within the subspace of
${\mathbb Q}^k$ defined by $l_1$
(because $q$ does not satisfy $l_1$, and neither does $qx$).
Hence, $L$ intersects this subspace in at most one point.
Because $p$ and hence also $p x \in L$ satisfies $l_1$,
we have thus shown that $(z_1,\dots,z_k)$ equals $p x$.
Since $p \neq q$ we conclude
that $y=0$ by the equations in the second line of the formula.
Now, consider the case that $l_2$ is satisfied.
Similarly as in the last case, $L$ intersects the subspace 
defined by $l_2$ in at most one point. 
Because $q \in L$ satisfies $l_2$, we have shown that
$(z_1,\dots,z_k)$ equals $q$.
The equations in the second line of the formula then imply that $x=y$.

Finally, we prove that
$S_{\mathbb Q}(u,v,w)$ has the following
pp-definition in $({\mathbb Q}; +,T)$:
\begin{align*}
\exists x,y. \; & x + v = w \; \wedge \; y + v = u \; \wedge \; T(x,y).
\end{align*}
Suppose first that $(u,v,w) \in S_{\mathbb Q}$.
Note that $x = w-v$ is not equal to $0$ because $v \neq w$.
If $u=v$, then $y=0$, and if $u=w$, then $x=w-v=u-v=y$ so
$T(x,y)$ is satisfied. 

Conversely, suppose that $(x,y) \in {\mathbb Q}^2$ satisfies the pp-formula above. The formula $T(x,y)$ implies that
$x \neq 0$ and hence $w \neq v$. Moreover, $T(x,y)$
implies that $y=0$ or $x=y$. If $y=0$, then $u=v$ and 
$(u,v,w) \in S_{\mathbb Q}$. If $x=y$, then $w-v=u-v$ and hence
$u=w$. Again $(u,v,w)$ is in $S_{\mathbb Q}$. 
\end{proof}

\subsection{Classification Result} \label{sect:cc1}

We will now use Lemma~\ref{lem:hard} in order to prove the
following definability result.

\begin{theorem}\label{thm:q}
Let
$\Gamma$ be first-order expansion of $({\mathbb Q}; +)$.
Then, either 
\begin{itemize}
\item each relation in $\Gamma$ 
has a quantifier-free Horn definition in $({\mathbb Q}; +)$, or
\item every non-empty relation of $\Gamma$ contains a tuple of the form $(0,\dots,0)$, or 
\item every first-order definable relation in $({\mathbb Q}; +)$ has a 
pp-definition in $\Gamma$.
\end{itemize}
\end{theorem}
\begin{proof}
Suppose that there is a non-empty $k$-ary relation $R$ of $\Gamma$ that
does not contain the tuple $(0,\dots,0)$. Then the $(k+1)$-ary relation
$R'(x_1,\dots,x_{k+1})$ defined by $R(x_1,\dots,x_k) \wedge x_{k+1}=0$ is non-empty,
and the relation defined by $R'(x,\dots,x)$ is empty. So we can apply
Proposition~\ref{prop:neq} and find that $\neq$ is pp-definable
in $(\mathbb Q; +,R')$ and hence also in $\Gamma$.
So assume in the following without loss of generality that $\Gamma$
contains the relation $\neq$.

Suppose that one of the relations of $\Gamma$ does not have a quantifier-free Horn definition in $({\mathbb Q}; +)$. 
Lemma~\ref{lem:hard} implies that the relation
$S_{\mathbb Q}$ has a pp-definition in $\Gamma$, and
Corollary~\ref{s-prop}
implies that every relation with a first-order definition
in $({\mathbb Q}; =)$ has a pp-definition in $\Gamma$.

Let $R$ be a relation with a first-order definition $\phi$ in $({\mathbb Q}; +)$. To find a pp-definition for $R$ in $\Gamma$,
we introduce a variable $u$ for 
every atomic formula of the form $x+y=z$ in $\phi$.
For each atomic formula $\psi$ in $\phi$ of the form
$x+y=z$, we replace $\psi$ by $u_\psi=z$ for a new variable $u_\psi$. The resulting formula consists of a boolean combination
of atomic formulas of the form $x=y$, which we know has a pp-definition $\phi'$ in $\Gamma$. For each
atomic formula $\psi$ in $\phi$ we add the conjunct 
$x+y=u_\psi$ to $\phi'$, and finally existentially quantify over all new variables.
It is straightforward to verify that the resulting formula is
a pp-definition of $R$ in $\Gamma$. 
\end{proof}

Theorem~\ref{thm:q} has immediate consequences for the
computational complexity of constraint satisfaction.

\begin{corollary}\label{cor:q}
Let $\Gamma$ be a structure with a finite relational signature and 
a first-order
definition in $(\mathbb Q; +)$ that contains the relation 
$\{(x,y,z) \; | \; x+y=z \}$. Then CSP$(\Gamma)$
is in P if all relations in $\Gamma$ have a quantifier-free Horn
definition over $(\mathbb Q; +)$, or if all non-empty relations contain a tuple of the form $(0,\dots,0)$, and is NP-hard otherwise. 
\end{corollary}
\begin{proof}
If all relations in $\Gamma$ have a quantifier-free Horn
definition over $(\mathbb Q;+)$, then Proposition~\ref{prop:tract}
implies that CSP$(\Gamma)$ is in P.
Otherwise, Theorem~\ref{thm:q} implies that in particular the
relation defined by $(x=y \wedge y \neq z) \vee (x \neq y \wedge y \neq z)$ is pp-definable in $\Gamma$. 
It follows from Theorem~\ref{thm:ecsp} that the constraint satisfaction problem
for this ternary relation is NP-hard. 
\end{proof}


\section{Affine Structures over the Rational Numbers}\label{sect:affine}

We will now consider affine additive structures over $\mathbb Q$.
The structure of this section is very similar to the structure of
Section~\ref{sect:q}: we begin by studying the definability of $\neq$ (Section~\ref{sect:neq2}) and of $S_D$ (in Section~\ref{sect:defsd})
and use these results to completely classify the problem in Section~\ref{sec:cc2}. The main proof in Section~\ref{sect:defsd}, however, is very different from the corresponding proof in Section~\ref{sect:defsq}.

Let us now formally define the problem at hand: define the operation
$f : {\mathbb Q}^3 \rightarrow {\mathbb Q}$ by
$f(a,b,c)=a-b+c$.
We study the constraint satisfaction problem for templates $\Gamma$
with a first-order definition in $(\mathbb Q; f)$ that 
contain the relation $\{(a,b,c,d) \; | \; a-b+c = d \}$.

\subsection{Definability of Disequality}\label{sect:neq2}

\begin{lemma}\label{lem:2trans}
Let $\Gamma$ be a structure with a first-order definition 
in $(\mathbb Q; f)$. Then there are at most four first-order definable binary relations: the empty relation, the full relation, the relation $\neq$,
and the relation $=$.
\end{lemma}
\begin{proof}
It suffices to show that Aut$(\Gamma)$
has precisely two orbits
on ${\mathbb R}^2$, namely 
$$O_1 = \{(x,x) \; | \; x \in \mathbb Q\} \quad \text{ and } \quad O_2 = \{(x,y) \; | \; x,y \in {\mathbb Q}, x \neq y \}\; .$$
These two orbits clearly partition ${\mathbb Q}^2$.
It is obvious that $O_1$ is an orbit, because for every $c \in \mathbb Q$ the mapping $x \mapsto x + c$
is an automorphism of $(\mathbb Q; f)$ and hence of $\Gamma$.
To see that $O_2$ is an orbit of pairs of reals, we apply linear interpolation: let $(a,b) \in O_2$ and $(c,d) \in O_2$ be arbitrary. The mapping $x \mapsto \frac{c-d}{a-b} (x - a) + c$ maps
$(a,b)$ to $(c,d)$ and it is an automorphism
of $(\mathbb Q; f)$, and hence of $\Gamma$. 
 \end{proof}
In the proof of Lemma~\ref{lem:2trans} we have in fact verified that
the automorphism group of $\Gamma$ is \emph{2-transitive},
i.e., that there is only one orbit of pairs of distinct elements with respect to the componentwise action of the automorphism group of $\Gamma$ on pairs.

\begin{theorem}[from~\cite{tcsps}]\label{thm:neq2}
Let $\Gamma$ be a relational structure with a 2-transitive automorphism group. If there is no pp-definition of $\neq$,
then there is an element $x$ of $\Gamma$ such that every non-empty relation in $\Gamma$ contains a tuple of
the form $(x,\dots,x)$.
\end{theorem}

\subsection{Definability of $S_D$} \label{sect:defsd}

The central step of the classification 
is the following result 
concerning pp-definability. 

\begin{lemma}\label{lem:indep2}
Let $\Gamma$ be a relational structure over an infinite domain $D$ 
such that $D^2$, $=$, $\neq$, and $\emptyset$
are the only pp-definable binary relations.
Suppose that $\Gamma$ contains a relation $Q$ such that there
are pairwise distinct $1 \leq i,j,k,l \leq n$ for which the following conditions hold:
\begin{enumerate}
\item $Q(x_1,\dots,x_n) \wedge x_i \neq x_j$ is satisfiable;
\item $Q(x_1,\dots,x_n) \wedge x_k \neq x_l$ is satisfiable;
\item $Q(x_1,\dots,x_n) \wedge x_i \neq x_j \wedge x_k \neq x_l$ is unsatisfiable.
\end{enumerate}
Then $S_D$
has a pp-definition in $\Gamma$.
\end{lemma}

We simplify the proof of Lemma~\ref{lem:indep2} by first
proving a slightly restricted version:

\begin{lemma}\label{lem:indep1}
Let $\Gamma$ be a relational structure over an infinite domain $D$
such that $D^2$, $=$, $\neq$, and $\emptyset$
are the only pp-definable binary relations.
Suppose that $\Gamma$ contains a relation $Q$ such that there
are $1 \leq i,j,k \leq n$ for which the following conditions hold:
\begin{enumerate}
\item $Q(x_1,\dots,x_n) \wedge x_i \neq x_j$ is satisfiable;
\item $Q(x_1,\dots,x_n) \wedge x_i \neq x_k$ is satisfiable;
\item $Q(x_1,\dots,x_n) \wedge x_i \neq x_j \wedge x_i \neq x_k$ is unsatisfiable.
\end{enumerate}
Then $S_D$
has a pp-definition in $\Gamma$.
\end{lemma}
\begin{proof}
The indices $i,j,k$ must be pairwise distinct, so
suppose for the sake of notation that $i=1$, $j=2$, $k=3$.
Consider the relation $R$ defined by 
$$ R(x_1,x_2,x_3) \equiv \exists x_4,\dots,x_n. Q(x_1,\dots,x_n) 
\wedge x_2 \neq x_3 \; .$$
We first note that $R$ is a non-empty relation: 
$Q(x_1,\dots,x_n)$ is satisfiable so the only way of
making $R$ empty is that every tuple $(s_1,\dots,s_n)$ in $Q$
satisfies $s_2=s_3$. This is impossible since we know that
there exists a tuple $(s_1,\dots,s_n) \in Q$ such that
$s_1 \neq s_2$. This implies  
$s_1 \neq s_3$ and contradicts the third condition.

Arbitrarily choose a domain element $a$. We first show
that there always exist elements $y,z \in D$ such that
$(a,y,z) \in R$. Let $A=\{a \in D \; | \; \exists y,z.R(a,y,z)\}$
and note that $A$ is
pp-definable. We know that $A$ is non-empty since $R$
is non-empty. 
Now assume that $A \subsetneq D$. First suppose that $|A|=1$. Then 
$$A'(x,y) \equiv A(x) \wedge A(y)$$ is non-empty and a strict subset of the equality relation, a contradiction.

If $|A|>1$, then consider
the pp-definable relation 
\[A'(x,y) \equiv A(x) \wedge A(y) \wedge x \neq y.\]
We see that $\emptyset \subsetneq A' \subsetneq \{(u,v) \in D^2\; | \; u \neq v\}$
which contradicts the fact that the only non-trivial
binary relations that are pp-definable from $\Gamma$ are
$=$ and $\neq$. Hence, $A=D$.

We now continue by considering the tuple $(a,y,z) \in R$.
By the third condition, we see that at least one of $y,z$
must equal $a$ in order to satisfy $R$. Let us consider the case
$R(a,a,z)$. Note that $(a,a,a) \not\in R$ due to the literal
$y \neq z$. We now show that $R(a,a,z)$ is satisfied
by any choice of $z$ except $a$. To see this, assume to the contrary
that there is a domain element $b \neq a$ such that $(a,a,b) \not\in R$.
Define $R'(x,z) \equiv R(x,x,z)$ and
note that $\emptyset \subsetneq R' \subsetneq  \{(u,v) \in D^2\; | \; u \neq v\}$ which contradicts the assumption that $=$ and $\neq $ are the only non-trivial pp-definable binary 
relations. Similarly, one can show that $R(a,y,a)$
holds for all $y \neq a$. Therefore $R=S_D$. 
\end{proof}

\begin{proof}[Proof of Lemma~\ref{lem:indep2}]
Assume for notational simplicity that $i=1,j=2,k=3$, and $l=4$. 
Define the $4$-ary relation $R$ by
$$R(x_1,x_2,x_3,x_4) \equiv \exists x_5,\dots,x_n. Q(x_1,\dots,x_n)$$
and consider the formula
$\phi = R(x,y,x',y') \wedge R(z',y',z,y) \wedge x' \neq z'.$
We claim that $\phi \wedge x \neq y$ and
$\phi \wedge y \neq z$ are satisfiable while
$\phi \wedge x \neq y \wedge y \neq z$ is not satisfiable. 
Then we can apply Lemma~\ref{lem:indep1} and are done.
First we make an observation:

\medskip

\noindent
\underline{Observation 1.} 
Define relation $R_1$ such that

\[R_1(u,v) \equiv \exists x,y. R(x,y,u,v) \wedge x \neq y.\]

\noindent
We know that $R(x,y,u,v) \wedge x \neq y$ is satisfiable so
$R_1$ is a non-empty relation.
Since $R_1(u,v) \wedge u \neq v$ is not satisfiable, we conclude
that $R_1$ is a non-empty subset of the equality relation. Consequently,
$R_1$ is the equality relation.
Analogously, define $R_2$ such that

\[R_2(u,v) \equiv \exists z,y. R(u,v,z,y) \wedge z \neq y\]

\noindent
and note that $R_2$ is the equality relation, too.
\medskip

\noindent
We now prove that
$\phi \wedge x \neq y \wedge y \neq z$ is not satisfiable. 
By using Observation 1, it follows that any solution $s$
satisfies
$x' = y'$ and $y'=z'$ --- this is impossible due to
the clause $x' \neq z'$.

Next, we prove that $\phi \wedge x \neq y$ is satisfiable; the case
$\phi \wedge y \neq z$ is symmetric.
Consider the relation 
\[U(u,v) \equiv \exists w. R(w,u,v,v) \wedge w \neq u.\]
\noindent
By the conditions on $R$, we know that $U$ is non-empty. Since $U$
is binary, we also know that $U$ either is the equality relation, the disequality relation, or
the full relation. We conclude
that $U$ is non-empty and symmetric.

By Observation 1, the clause $x \neq y$ 
has the effect that
every solution $s$ must satisfy $x'=y'$. The solution also has to satisfy
$x' \neq z'$ which implies that $y' \neq z'$. Observation 1
now tells us that $z=y$ and we conclude that
every solution 
satisfies $x'=y'$ and $z=y$. 
We define 
\[\phi'=R(x,y,x',x') \wedge R(z',y',z,z) \wedge x' \neq z' \wedge x \neq y\]
\noindent
Thus, $\phi'$ is
satisfiable if and only if $\phi \wedge x \neq y$ is satisfiable.
We will now construct a concrete
satisfying assignment $s$ to the variables of $\phi'$.

Arbitrarily choose a tuple $(a,b) \in U$ and let $s(y)=a$, $s(x')=b$.
By the conditions on $U$, there exists an element $c$ such that
$(c,a,b,b) \in R$ and $c \neq a$; we let $s(x)=c$. 
Furthermore, we know that $s(x')=s(y')$ and $s(z)=s(y)$ so
$s(y')=b$ and $s(z)=a$.
At this point, we see that the assignment
$s$ satisfies the clauses $R(x,y,x',x')$ and $x \neq y$.

We know that $(a,b) \in U$ so $(b,a) \in U$, too, and
there exists a value $d$ such that $(d,b,a,a) \in R$ and $d \neq b$.
Now, let $s(z')=d$ and note that $R(z',y',z,z)$ is satisfied by $s$.
Finally, $s(x')=b \neq d = s(z')$ so the clause
$x' \neq z'$ is satisfied and the proof is completed. 
\end{proof}

\subsection{Classification Result} \label{sec:cc2}
We are now ready to prove the classification result for the affine case.

\begin{theorem}\label{thm:affine}
Let $\Gamma$ be a first-order expansion of $(\mathbb Q; f)$.
Then, either 
\begin{itemize}
\item each relation in $\Gamma$ has a quantifier-free Horn definition
in $(\mathbb Q; f)$, or 
\item every non-empty relation of $\Gamma$ contains a tuple of the form $(0,\dots,0)$, or
\item every first-order definable relation
in $(\mathbb Q; f)$ has a pp-definition in $\Gamma$.
\end{itemize}
\end{theorem}
\begin{proof}
Suppose that there is a non-empty $k$-ary relation 
$R$ of $\Gamma$ that
does not contain the tuple $(0,\dots,0)$. The proof of 
Lemma~\ref{lem:2trans} shows that $\Gamma$ is 2-transitive,
and hence by the contraposition of Theorem~\ref{thm:neq2} the relation $\neq$ is pp-definable.
So assume in the following without loss of generality that $\Gamma$
contains the relation $\neq$.

Let $R$ be a relation in $\Gamma$ that does not have a quantifier-free
Horn definition in $({\mathbb Q}; f)$.
Let $\phi(x_1,\dots,x_n)$ be a reduced definition of $R$ in $({\mathbb Q}; f)$ (see Section~\ref{sect:q}). 
Then there must be a clause $C$ in $\phi$ 
with at least two positive literals 
$f(x_{i_1},x_{i_2},x_{i_3})=x_{i_4}$ and $f(x_{j_1},x_{j_2},x_{j_3})=x_{j_4}$.
Let $Q(x_1,\dots,x_n,x_{n+1},x_{n+2})$ be the relation defined by

\[ \begin{array}{l}\phi(x_1,\dots,x_n) \; \wedge \;  \bigwedge_{l \in C \setminus \{l_1,l_2\}} \neg l \; \wedge \\ x_{n+1} = f(x_{i_1},x_{i_2},x_{i_3}) \; \wedge \; x_{n+2} =  f(x_{j_1},x_{j_2},x_{j_3}). \end{array} \]

This relation $Q$ is clearly pp-definable over $({\mathbb Q}; R,f,\neq)$. 
We claim that $Q$ satisfies the conditions of Lemma~\ref{lem:indep2}
(which is applicable due to Lemma~\ref{lem:2trans})
with respect
to the arguments indexed by $i_4,n+1,j_4$, and $n+2$
(or the conditions of Lemma~\ref{lem:indep1} if $i_4=j_4$; this
remark also applies to all other places where we appeal to Lemma~\ref{lem:indep2}).
Since $\phi$ is reduced, there is a tuple $t \in R$ that satisfies
$l_2$ and does not satisfy all other literals in $\phi$.
Now, the extended tuple $t_1 = (t[1],\dots,t[n],t[i_4],t[j_4])$ clearly satisfies $Q$, and we have $t_1[i_4] \neq t_1[n+2]$ as required 
in the conditions for Lemma~\ref{lem:indep1}. 
There is also a tuple $t_2 \in R$ that
satisfies $l_1$ and does not satisfy all other literals in $C$, and we can argue similarly to find a second tuple showing the second condition of 
Lemma~\ref{lem:indep2}.

Finally, suppose for contradiction that there is a 
tuple $t_3$ in $Q$ where $t_3[i_4] \neq t_3[n+1]$
and $t_3[j_4] \neq t_3[n+2]$. Because this tuple satisfies
in particular the clause $C$ from $\phi$, the
conjunct $\bigwedge_{l \in C \setminus \{l_1,l_2\}} \neg l$
implies that either $l_1$ or $l_2$ is satisfied. 
But then the equalites 
$x_{n+1} = f(x_{i_1},x_{i_2},x_{i_3})$ and 
$x_{n+2} =  f(x_{j_1},x_{j_2},x_{j_3})$ imply that 
$t_3[i_4] = t_3[n+1]$ or $t_3[j_4] = t_3[n+2]$, a contradiction.
Hence, Lemma~\ref{lem:indep1} applies, $S_{\mathbb R}$ is
pp-definable over $({\mathbb Q}; Q)$ and therefore also over $({\mathbb Q}; R,f,\neq)$ and $\Gamma$. The result follows from Corollary~\ref{s-prop}.
 \end{proof}

The next corollary is a direct consequence of
Proposition~\ref{prop:tract},
Theorem~\ref{thm:affine}, and Corollary~\ref{s-prop}.

\begin{corollary}
Let $\Gamma$ be an expansion of $(\mathbb Q; \{(a,b,c,d) \; | \; a-b+c=d\})$ by finitely
many first-order definable relations.
If each relation in $\Gamma$ 
has a quantifier-free Horn definition
in $(\mathbb R,f)$, or if each non-empty relation contains a tuple of the form $(0,\dots,0)$,
then CSP$(\Gamma)$ is in P. Otherwise,
CSP$(\Gamma)$ is NP-hard.
\end{corollary}

\section{Concluding Remarks}
\label{sect:conc}

We have presented classification results for certain algebraic
constraint satisfaction problems, and the results are to a large extent
based on dichotomy results for logical definability.
We feel that
the results and ideas presented in this paper can be extended
in many different directions. Hence, it seems worthwhile to
provide some concrete suggestions for future work.

The results and proof techniques in Section~\ref{sect:q} appear to be generalisable 
to many different templates
defined over various structures.
One example is the natural and important class of structures that are
definable in {\em Presburger arithmetics}~\cite{Presburger}, 
i.e., structures that are first-order
definable over the integers with addition $(\mathbb Z;+)$.
We note that the following can be obtained by slightly modifying 
Corollary~\ref{cor:q}. 

\begin{corollary}
Let $\Gamma$ be a relational structure with a quantifier-free first-order
definition in $(\mathbb Z; +)$ that contains the relation 
$\{(x,y,z) \; | \; x+y=z \}$. Then CSP$(\Gamma)$
is in P if all relations in $\Gamma$ have a quantifier-free Horn
definition over $(\mathbb Z; +)$, or if all non-empty relations contain a tuple of the form $(0,\dots,0)$.  Otherwise, CSP$(\Gamma)$ is NP-hard. 
\end{corollary}

There is an important difference between this result and a full classification result: we have replaced
\emph{first-order definability} with
\emph{quantifier-free first-order definability} in the statement
of the result, and the reason is that
$({\mathbb Z}; +)$ does not admit quantifier elimination.
Is there still a complexity dichotomy if we look at the class of CSPs with an template that is first-order definable in $(\mathbb Z;+)$? 
This appears to be a difficult question.

The results presented in Section 5 have strong connections with earlier
work on the complexity of {\em disjunctive} constraints~\cite{BroxvallJonssonRenz,Disj}. 
We say that $\neq$ is \emph{1-independent} with respect to a $\tau$-structure
$\Gamma$ if and only if 
for every primitive positive $\tau$-formula $\phi$ with free variables $x,y,z,w$
the following holds: if $\phi \wedge x \neq y$ and $\phi \wedge z \neq w$ are satisfiable, then
so is $\phi \wedge x \neq y \wedge z \neq w$.
Assume that CSP$(\Gamma)$ is tractable and let
$\Gamma'$ denote the set of all relations that can be defined
by (quantifier-free) conjunctions of disjunctions over $\Gamma$ containing at most one literal that is not of the form $x \neq y$. The following has been
shown in~\cite{BroxvallJonssonRenz,Disj}; it does 
not imply our result since it only makes a
statement about a constraint language $\Gamma'$ of the form described above.

\begin{theorem}[from \cite{BroxvallJonssonRenz,Disj}]
Let $\Gamma$ and $\Gamma'$ be defined as above, and assume that P $\neq$ NP. Then CSP$(\Gamma')$ is tractable
if and only if $\neq$ is 1-independent with respect to $\Gamma$.
\end{theorem}

We have already mentioned that 
the structures studied in this paper are in general not 
$\omega$-categorical. 
However, torsion-free divisible abelian groups such as $(\mathbb Q; +)$ and all structures first-order definable in such groups are strongly
minimal (see e.g.~Corollary 3.1.11 in~\cite{Marker}), and hence
categorical in all uncountable cardinals. 
This is interesting from a constraint satisfaction point of view because
of the following preservation theorem.

\begin{theorem}[of~\cite{BodHilsMartin}]\label{thm:galois}
Let $\Gamma$ be an uncountably categorical structure with a countable relational signature and an uncountable domain.
Then a first-order definable 
relation $R$ has a pp-definition in $\Gamma$
if and only if $R$ is preserved by all \emph{infinitary} polymorphisms of $\Gamma$.
\end{theorem}

Note that this theorem is weaker than the corresponding
theorem for $\omega$-categorical structures~\cite{BodirskyNesetrilJLC}, because we have to assume that $R$ is first-order definable, and that $R$ is not only preserved by the finitary, but also by the infinitary polymorphisms of $\Gamma$.
Since our classification result is purely in terms of primitive positive definability of first-order definable relations, it is an interesting 
question to describe the polymorphisms that guarantee tractability
for structures $\Gamma$ 
with a first-order definition in $(\mathbb Q;+)$
(Theorem~\ref{thm:galois} shows that such polymorphisms do exist).


\section*{Acknowledgements} 
We want to thank Barnaby Martin for comments on an earlier version of the paper.
Peter Jonsson is partially supported by the {\em Center for Industrial Information Technology}
({\sc Ceniit}) under grant 04.01 and by the {\em Swedish Research Council} (VR) under
grant 2006-4532.



\end{document}